\documentclass[twocolumn,secnumarabic,showpacs,amssymb, amsmath,tightenlines,aps,prl]{revtex4}
\usepackage{longtable}
\usepackage{bm}
\usepackage{epsfig}
\usepackage{graphicx}\usepackage{rotating}

\begin{document}

\title[Critical Scaling in the Weak Itinerant Ferromagnet UIr]
{Critical Scaling of the Magnetization and Magnetostriction in the
Weak Itinerant Ferromagnet UIr}

\author{W. Knafo$^{1,2,3}$, C.
Meingast$^{1}$, S. Sakarya$^{4}$, N.H. van Dijk$^{4}$,\\ Y.
Huang$^{5}$, H. Rakoto $^{3}$, J.-M. Broto $^{3}$, and H. v.
L\"{o}hneysen$^{1,2}$}

\address{$^{1}$ Forschungszentrum Karlsruhe, Institut f\"{u}r Festk\"{o}rperphysik, D-76021 Karlsruhe, Germany\\
$^{2}$ Physikalisches Institut, Universit\"{a}t Karlsruhe, D-76128
Karlsruhe, Germany\\
$^{3}$ Laboratoire National des Champs Magn\'{e}tiques Puls\'{e}s,
143 Avenue de Rangueil, 31400 Toulouse, Cedex 4, France\\
$^{4}$ Fundamental Aspects of Materials and Energy, Delft
University of Technology, Mekelweg 15, 2629 JB Delft, The
Netherlands\\
$^{5}$ Van der Waals - Zeeman Institute, University of Amsterdam,
Valckenierstraat 65, 1018 XE Amsterdam, The Netherlands}

\date{\today}

\begin{abstract}

The weak itinerant ferromagnet UIr is studied by magnetization and
magnetostriction measurements. Critical behavior, which
surprisingly extends up to several Tesla, is observed at the Curie
temperature $T_C\simeq45$ K and is analyzed using Arrott and
Maxwell relations. Critical exponents are found that do not match
with any of the well-known universality classes. The
low-temperature magnetization $M_s\simeq0.5$ $\mu_B \cong const.$
below 3 T rises towards higher fields and converges
asymptotically around 50 T with the magnetization at $T_C$. From
the magnetostriction and magnetization data, we extract the
uniaxial pressure dependences of $T_C$, using a new method
presented here, and of $M_s$. These results should serve as a
basis for understanding spin fluctuations in anisotropic
itinerant ferromagnets.

\end{abstract}


\maketitle

The fact that a superconducting pocket develops at the quantum
critical point (QCP) of several heavy-fermion systems indicates
that magnetic fluctuations, which are enhanced at the QCP, may
play an important role for forming the Cooper pairs
\cite{sarrao07,flouquet06,thalmeier05}, suggesting that
superconductivity might be mediated by these fluctuations
\cite{mathur98}, instead of phonons as in conventional
superconductors. The recent observation of magnetic-field induced
superconductivity in URhGe confirmed that magnetism plays a
central role for the superconducting properties of such systems
\cite{levy05}. The issue of understanding the interplay between
magnetism and superconductivity is not restricted to heavy-fermion
physics, since a magnetic pairing mechanism might be operative for
the cuprate superconductors as well
\cite{moriya00,storey07,sarrao07}. A precise characterization of
the properties of the magnetically ordered state is, thus, an
important first step in understanding how superconductivity can
develop in correlated electron systems.

The weak itinerant ferromagnet UIr has recently attracted
attention because it becomes superconducting under pressure
\cite{akazawa04}. Further, it is the second known case (after
CePt$_3$Si \cite{bauer04}) of superconductivity in a
non-centrosymmetric crystal with strong electronic correlations.
UIr orders ferromagnetically below $T_C\simeq46$ K, its
low-temperature saturated moment $M_s\simeq0.5$ $\mu_B$/U-atom
being much smaller than the effective moment $M_{eff}\simeq3.6$
$\mu_B$/U-atom extracted from the high temperature susceptibility
\cite{dommann87,galatanu05}. The small ratio $M_s/M_{eff}=0.14$
is probably due to the combination of strong longitudinal magnetic
fluctuations, consequence of the itinerant character of the 5$f$
electrons \cite{moriya85}, and of crystal-field effects
\cite{galatanu05,sakarya07}. Hydrostatic pressure destabilizes
ferromagnetism and leads to a QCP at $p_C\simeq24$ kbar, in the
vicinity of which superconductivity develops with a maximum
critical temperature $T_{sc}^{max}\simeq0.14$ K \cite{akazawa04}.
As summarized in Table \ref{tableintro}, UIr has strong
similarities with the itinerant ferromagnets UGe$_2$ and URhGe,
which become superconducting under pressure \cite{saxena00} and
at ambient pressure \cite{aoki01}, respectively, and, to a lesser
degree, with ZrZn$_2$, which is an archetype of itinerant
ferromagnetism, without superconductivity
\cite{uhlarz04,yelland05}.

\begin{table}[b]
\caption{Comparison of $T_C$, $M_s$, $M_s/M_{eff}$, $p_C$, and
$T_{sc}^{max}$ for the itinerant ferromagnets UIr, UGe$_2$, URhGe,
and ZrZn$_2$.}
\begin{ruledtabular}
\begin{tabular}{lcccc}
&UIr&UGe$_2$&URhGe&ZrZn$_2$\\[-1ex]
&\footnotesize\cite{akazawa04,dommann87,galatanu05} &\footnotesize\cite{saxena00,huxley01}&\footnotesize\cite{aoki01,hardy05}&\footnotesize\cite{uhlarz04,yelland05}\normalsize\\
\hline
$T_C$ (K)&45&53&9.5&28.5\\
$M_s$ ($\mu_B$)&0.5&1.5&0.4&0.17\\
$M_s/M_{eff}$&0.14&0.43&0.11&0.09\\
$p_C$ (kbar)&24&16.5&($<$ 0)&16.5\\
$T_{sc}^{max}$ (K)&0.14&0.80&\;\,0.25$^{\star}$&-\\
\end{tabular}
\end{ruledtabular}
\label{tableintro} \flushleft \small $^{\star}$: in URhGe,
reentrant superconductivity is associated with
$T_{sc}^{max}\simeq0.4$ K at $\mu_0H\simeq12$ T
($\mathbf{H}\parallel\mathbf{b}$) \cite{levy05}. \normalsize
\end{table}

In this Letter, we present a study of the magnetic properties of
UIr. Power laws, characteristics of critical ferromagnetic
fluctuations, are observed at $T_C$ in the magnetization $M(H)$
and, for the first time in a ferromagnet, in the magnetostriction
$\lambda(H)$. Surprisingly, this critical regime extends to
fields of several Tesla. A description of the critical
magnetization and magnetostriction is made using Arrott and
Maxwell relations, from which we derive a new method to extract
the uniaxial pressure dependences of $T_C$. The pressure
dependences of $M_s$ are also extracted. These results confirm
the anisotropic nature of magnetism in UIr \cite{galatanu05}. A
connection is made with Moriya-like spin-fluctuation theories
\cite{moriya85,moriya95,takahashi06}.

\begin{figure}[b]
    \centering
    \epsfig{file=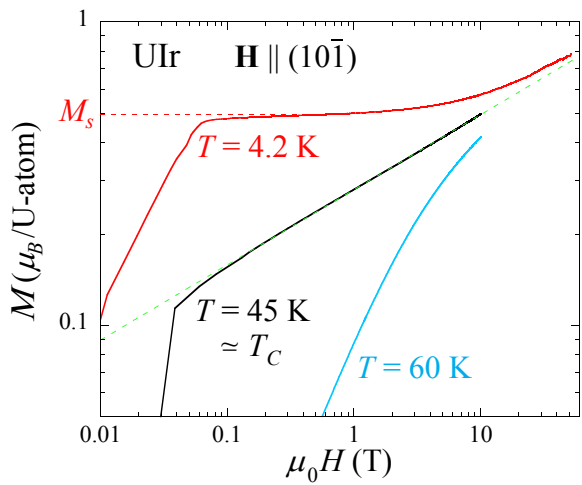,width=76mm}
    \caption{(color online) Magnetization versus field of UIr at $T=4$,
    45, and 60 K, with $\mathbf{H}\parallel$ (10$\overline{1}$). The dashed line corresponds to the power law at $T_C$.}
    \label{magnetization}
\end{figure}

\begin{figure}[t]
    \centering
    \epsfig{file=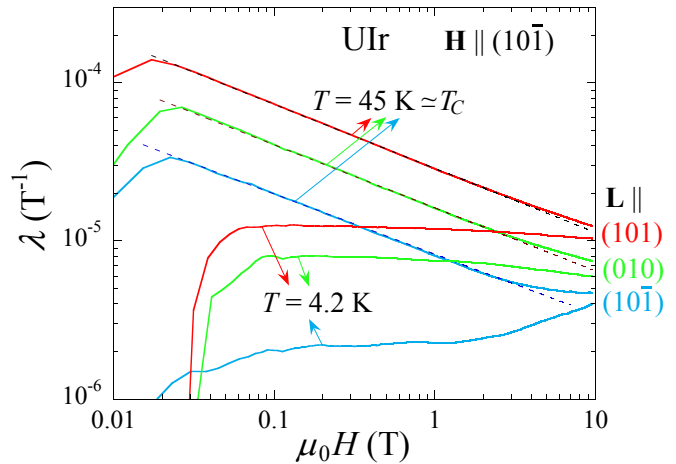,width=87mm}
    \caption{(color online) Magnetostriction versus field of UIr, measured at $T=4$ and 45
    K for $\mathbf{L}\parallel$ (10$\overline{1}$), (101), and (010), with $\mathbf{H}\parallel$(10$\overline{1}$).
    The dashed lines correspond to the fits of the data at $T_C$ by power laws.}
    \label{magnetostriction}
\end{figure}

Single crystals of UIr were grown by the Czochralski technique in
a tri-arc furnace. The magnetization up to 10 T was measured at
several temperatures using a VSM insert in a PPMS from Quantum
Design, with field sweep rates of 1 mT/s. Magnetization was also
measured at $T=4.2$ K using a pulsed field up to 52 T at the
Laboratoire National des Champs Magn\'{e}tiques Puls\'{e}s in
Toulouse, with a pulse rise of 40 ms (see also \cite{sakarya07}).
Magnetostriction was measured using a high-resolution capacitive
dilatometer \cite{meingast90,pott83} with field sweep rates of
0.5 T/min. The cell is rotatable, allowing both longitudinal and
transverse measurements of the length $L$, which was measured
along the (10$\overline{1}$), (101), and (010) axes of UIr. In all
measurements, the magnetic field $\mathbf{H}$ was applied along
the easy (10$\overline{1}$) axis.

In Fig. \ref{magnetization}, the magnetization of UIr is shown in
a $M$ versus $H$ log-log plot, at $T=4$, 45 ($\simeq T_C$), and
60 K and for $\mathbf{H}\parallel$ (10$\overline{1}$). At 4 K,
after a domain alignment achieved at 0.04 T, $M(H)$ is nearly
constant and equals $M_s\simeq0.5$ $\mu_B$/U-atom for
$0.04<\mu_0H\lesssim3$ T. At $T_C$, $M(H)$ follows a power law
$M\propto H^{1/\delta}$, with $\delta\simeq4.04\pm0.05$,
indicative of a critical regime up to the highest measured field
of 10 T. The magnetization is also shown in the paramagnetic
regime, at 60 K. Above 2 T, the $M(H)$ data at 4 K increase
significantly and, quite surprisingly, appear to merge with the
high-field extrapolation of the power law (dashed line) above
roughly 50 T. Even around 50 T, $M(H)$ continues to increase,
being still a factor of roughly 5 smaller than the effective
moment. The above behavior contrasts strongly with the
magnetization curves of localized ferromagnetic systems such as
EuS, for which $M(H)$ at $T_C$ only shows a power law at
relatively small fields, and saturates at high fields to a
field-independent saturated moment $M_s$ \cite{siratori82}. We
attribute the unusually large field range of the power-law
behavior at $T_C$ and of the field-induced increase of $M(H)$ at
4 K to the strong itinerant character of UIr. In this picture, the
magnetization increases above 2 T because the magnetic field
quenches the longitudinal fluctuations.

In Fig. \ref{magnetostriction}, the magnetostriction of UIr,
defined by:
\begin{eqnarray}
\lambda_i=\frac{1}{L_i}\frac{\partial L_i}{\partial
(\mu_0H)}=-\frac{\partial M}{\partial p_i},
    \label{maxwellmagnetostriction}
\end{eqnarray}
where $p_i$ the uniaxial stress applied along $i$, is shown at 4
and 45 K ($\simeq T_C$) in a $\lambda$ versus $H$ log-log plot,
for $\mathbf{L}\parallel$ (10$\overline{1}$), (101), and (010)
and $\mathbf{H}\parallel$ (10$\overline{1}$). At 4 K and for
$0.04<\mu_0H\lesssim3$ T, the coefficients $\lambda_i(H)$ are
roughly constant. For $0.03<\mu_0H\lesssim3$ T, critical power
laws are, for the first time in a ferromagnet, observed in the
magnetostriction at $T_C$, which varies as $\lambda_i\propto
H^{-1/\delta_i'}$ with $\delta_i'\simeq$ 2.63, 2.47, and
$2.51\pm0.03$, for $i=$ (10$\overline{1}$), (101), and (010),
respectively. Above 3 T, deviations from the critical regime at
$T_C$, but also from the constant-$\lambda$ regime at 4 K, are
observed, indicating a crossover towards a high-field regime. The
crossover in $\lambda_i(H)$ at 4 K and $T_C$, as well as the one
in $M(H)$ at 4 K, could be associated with the quenching of the
longitudinal magnetic fluctuations. However, and contrary to the
magnetostriction, the magnetization at $T_C$ shows no apparent
crossover between a "low-field" critical regime and a "high-field"
quenching of the longitudinal fluctuations.

\begin{table}[t] \caption{Critical exponents
$\beta$, $\gamma$, $\delta$, and $\delta'$ from the fit of the
magnetization and for different universality classes
\cite{collins89}.}
\begin{ruledtabular}
\begin{tabular}{lcccc}
&$\beta$&$\gamma$&$\delta$&$\delta'$\\
\hline
UIr (best fit) &0.355&1.07&4.01&2.20\\
3D Heisenberg&0.367&1.388&4.78&2.77\\
3D XY&0.345&1.316&4.81&2.54\\
3D Ising&0.326&1.238&4.80&2.32\\
2D Ising&0.125&1.75&15&2.14\\
Mean Field&0.5&1&3&3
\end{tabular}
\end{ruledtabular}
\label{tablearrott}
\end{table}

\begin{figure}[t]
    \centering
    \epsfig{file=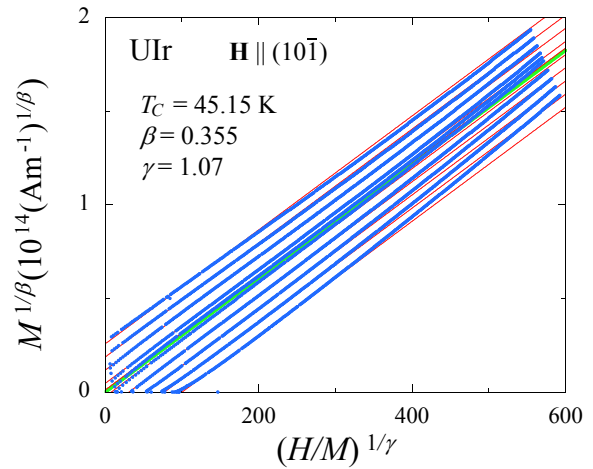,width=76mm}
    \caption{(color online) Arrott plot of magnetic isotherms for $41.5\leq T\leq49.5$ K. Thin red lines
    show fits of Eq. (\ref{arrottequation}) to the data, thick green line denotes the isotherm at $T_C$.}
    \label{arrottplot}
\end{figure}

In the following, we characterize the critical properties of the
magnetization and magnetostriction close to $T_C$, using the
Arrott equation of state \cite{arrott67}:
\begin{eqnarray}
M^{1/\beta}=c_1\left(\frac{H}{M}\right)^{1/\gamma}-c_2\left(T-T_C\right).
    \label{arrottequation}
\end{eqnarray}
Magnetization was measured up to 10 T at nine temperatures between
41.5 and 49.5 K. A best fit of Eq. (2) to the $M(H,T)$ data for
$42.5\leq T\leq47.5$ K and $0.2<\mu_0H<3$ T yields
$T_C=45.15\pm0.2$ K, $\beta=0.355\pm0.05$, $\gamma=1.07\pm0.05$,
$c_1=3.0\times10^{11}$ (A/m)$^{1/\beta}$, and
$c_2=7.0\times10^{12}$ (A/m)$^{1/\beta}$K$^{-1}$. Fig.
\ref{arrottplot} shows the resulting plot of $M^{1/\beta}$ versus
$(H/M)^{1/\gamma}$. Excellent fits are obtained in the whole $T$
range, with slight deviations for $\mu_0H\gtrsim5$ T. At $T=T_C$,
Eq. (\ref{arrottequation}) gives:
\begin{eqnarray}
M(H,T_C)= c_1^{\gamma/\delta} H^{1/\delta},
    \label{magnTC}
\end{eqnarray}
with $\delta=(\beta+\gamma)/\beta$. Using the exponents $\beta$
and $\gamma$ determined above, we obtain $\delta=4.01$, which
agrees very well with the value of $\delta=4.04$ from the fit of
$M(H)$ at 45 K. As shown in Table \ref{tablearrott}, the exponents
$\beta$, $\gamma$, and $\delta$ obtained for UIr do not belong to
any of the well-known universality classes \cite{collins89}. This
raises the question whether the critical regime of UIr is
non-universal or if it belongs to an unknown universality class.
Theoretical support, as well as an experimental determination of
the magnetic exchange, are needed to describe these unusual
critical properties. The presence of strong longitudinal magnetic
fluctuations could possibly play a role in the critical regime of
UIr.

The $H$-power law of $\lambda_i$ at $T_C$ is easily derived by
taking the $p_i$ derivative of Eq. (\ref{arrottequation}) and by
assuming that $\partial T_C/\partial p_i$ is the relevant pressure
derivative (i.e., $\partial c_1/\partial p_i=0$), which gives
\begin{eqnarray}
&\lambda_i(T_C)=-{\displaystyle A\frac{\partial T_C}{\partial
p_i}H^{-1/\delta'}},
   \label{Dershort}
\end{eqnarray}
where $A=c_1^{-\gamma/\delta'}c_2\gamma/\delta$ and
$\delta'=(\beta+\gamma)/(1-\beta)$. Using the values of $\beta$
and $\gamma$ determined above, we obtain the exponent
$\delta'=2.20$, which agrees, within 15 $\%$, with the exponents
$\delta_i'\simeq$ 2.63, 2.47, and $2.51$ obtained from the fit of
$\lambda_i(H)$ at $T_C$ for $i=$ (10$\overline{1}$), (101), and
(010), respectively. To our knowledge, this is the first time that
the isothermal magnetostriction of a ferromagnet at $T_C$ is
expressed as a power law of $H$, in function of the critical
exponents $\beta$ and $\gamma$ \cite{belov56}. Further, Eq.
(\ref{Dershort}) directly yields $\partial T_{C}/\partial
p_i=-0.24\pm0.05$, $-0.84\pm0.1$, and $-0.48\pm0.05$ K/kbar, for
$i=$ (10$\overline{1}$), (101), and (010), respectively, $A$ and
$\delta'$ = ($\beta+\gamma)(1-\beta)$ being extracted from the fit
of $M(H,T)$. These results agree perfectly with those obtained in
Ref. \cite{sakarya08} using the Ehrenfest relation:
\begin{eqnarray}
\frac{\partial T_C}{\partial
p_i}=\frac{\Delta\alpha_iV_mT_C}{\Delta c_p},
    \label{dTcdpehrenfest}
\end{eqnarray}
where $\Delta\alpha_i$ and $\Delta c_p$ are the jumps in the
thermal expansivity and specific heat at $T_C$, respectively, and
$V_m=2.45\times10^{-5}$ m$^3$/mol is the molar volume
\cite{dommann87} (Table \ref{tablepressdep}). The sum of the three
uniaxial pressure dependences leads to the hydrostatic pressure
dependence $\partial T_{C}/\partial p_h=-1.6\pm0.2$ K/kbar
\cite{note_volume}, which agrees well with the variation of $T_C$
under hydrostatic pressure \cite{akazawa04}.

\begin{table}[b]
\caption{Uniaxial and hydrostatic pressure dependences of $T_C$
and $M_s$ of UIr, and their normalized ratio $\rho_i$.}
\begin{ruledtabular}
\begin{tabular}{lcccc}
\;\;\;\;\;\;\;\;\;\;\;\;\;\;\;\;\;\;\;\;\;\;\;\;\;\;\;\;\;\;\;\;\;\;\;\;\;\;\;\;\;\;\;\;\;\;$\mathbf{p}\parallel$&(10$\overline{1}$)&(101)&(010)&$p_h$\\
\hline
$\partial T_{C}/\partial p_i$ {\footnotesize(K/kbar)} &&&&\\[-1ex]
{\footnotesize from Eq. (\ref{Dershort})} &-0.24&-0.84&-0.48&-1.6\\[-1ex]
{\footnotesize from Eq. (\ref{dTcdpehrenfest}) (Ehrenfest relation) \cite{sakarya08}}&-0.22&-0.88&-0.48&-1.6\\[1.5ex]
$\partial M_s/\partial p_i$ {\footnotesize(10$^{-3}$ $\mu_B$/kbar)} &-0.99&-5.2&-3.3&-9.5\\[0.5ex]
$\rho_i={\displaystyle\frac{(\partial \mathrm{ln} T_C/\partial
p_i)}{(\partial \mathrm{ln} M_s/\partial p_i)}}$&2.4&1.9&1.6&1.8\\
\end{tabular}
\end{ruledtabular}
\label{tablepressdep}
\end{table}

Using Eq. (\ref{maxwellmagnetostriction}), we extract the values
of $\partial M_s/\partial p_i=-\lambda_i(T=4\;\rm{K})$, with
$\lambda_i(T=4\;\rm{K})$ taken at $\mu_0H=1$ T for $i=$
(10$\overline{1}$), (101), or (010), from the low-temperature
magnetostriction (Fig. 2), see Table III. The sum over $\lambda$
yields the hydrostatic pressure dependence $\partial M_s/\partial
p_h=-9.5\times10^{-3}$ $\mu_B$/kbar \cite{note_volume}, which
agrees very well with measurements of $M_s$ under pressure
\cite{akazawa04}.

As seen from Table \ref{tablepressdep}, $T_C$ and $M_s$ are much
more sensitive to uniaxial pressures applied along the hard axes
(101) and (010) than to uniaxial pressures applied along the easy
axis (10$\overline{1}$). The strong anisotropy of these pressure
dependences (factor of up to five between different axes) results
presumably from single-ion, exchange, and/or hybridization
anisotropies. To compare these effects, we introduce the ratio:
\begin{eqnarray} \rho_i=\frac{(\partial\mathrm{ln}T_C/\partial p_i)}{(\partial\mathrm{ln}M_s/\partial
p_i)},
    \label{rho}
\end{eqnarray}
which is equal to 2.4, 1.9, 1.6, and 1.8 for $i=$
(10$\overline{1}$), (101), (010), and $h$ ($h$ = hydrostatic),
respectively (Table \ref{tablepressdep}). Not surprisingly,
$\rho_i$ is more isotropic than $\partial T_C/\partial p_i$ and
$\partial M_s/\partial p_i$, varying by about 50 \% from one axis
to another. In Moriya's theory of spin fluctuations
\cite{moriya85}, $T_C$ is given by:
\begin{eqnarray}
T_C\propto M^{3/2}_sT_A^{3/4}T^{1/4}_0,
    \label{Tc}
\end{eqnarray}
where $T_A$ and $T_0$ are two energy scales related to correlated
and uncorrelated spin fluctuations, respectively. From Eq.
(\ref{Tc}), Takahashi and Kanomata \cite{takahashi06} derived
\begin{eqnarray}
\frac{(\partial \mathrm{ln} T_C/\partial p)}{(\partial \mathrm{ln}
M_s/\partial p)}=\frac{3}{2}-\frac{2\gamma_{0,A}}{\gamma_m}.
    \label{rhotakahashi}
\end{eqnarray}
with $\gamma_{0,A}=(3\gamma_A+\gamma_0)/4$ and the Gr\"{u}neisen
parameters
$\gamma_A=-\partial\mathrm{ln}T_A/\partial\mathrm{ln}V$,
$\gamma_0=-\partial\mathrm{ln}T_0/\partial\mathrm{ln}V$, and
$\gamma_m=\partial\mathrm{ln}(M_s^2)/\partial$ln$V$. In such
models, the magnetic anisotropy is not considered and all the
formulas are isotropic. A straight-forward generalization of the
expressions of Takahashi and Kanomata to the anisotropic case of
UIr leads to the very anisotropic parameters
$\gamma_{0,A,i}/\gamma_{m,i}=-0.47$, -0.18, -0.06, and -0.17, for
$i=$ (10$\overline{1}$), (101), (010), and $h$, respectively. We
note that one could apply alternative models, such as that of
Kaiser and Fulde \cite{kaiser88}, to describe the magnetostriction
of itinerant ferromagnets.

In heavy-fermion systems, a stronger anisotropy seems to stabilize
superconductivity (compare cf. CeCoIn$_5$ and CeIn$_3$
\cite{sarrao07,mathur98,knafo03}). As proposed in Ref.
\cite{monthoux08}, anisotropy could also be of importance for the
development of superconductivity in U-based heavy-fermion systems.
Indeed, since it fixes the easy and hard magnetic axes, anisotropy
has a crucial influence on the magnetic fluctuations, and has to
be considered in the scenarios of magnetically mediated
superconductivity \cite{moriya00}. The anisotropy of the results
reported here, which depends on the quantity extracted
($\gamma_{0,A,i}/\gamma_{m,i}$, $\partial T_C/\partial p_i$, and
$\partial M_s/\partial p_i$ are more anisotropic than $\rho_i$),
suggests that the spin fluctuation theories
\cite{moriya85,takahashi06} should incorporate the magnetic
anisotropy for a better description of UIr.

In conclusion, we have studied the magnetic properties of the
itinerant ferromagnet UIr at ambient pressure. Special attention
was given to the critical regime at $T_C$, which surprisingly
persists up to several Tesla in the magnetization and
magnetostriction. A crossover to a high-field regime, where the
magnetic fluctuations are increasingly quenched by the field, was
observed above 3 T. In the high-field region $\mu_0 H \gtrapprox
50 T$, the low-temperature magnetization and the magnetization at
$T_C$ were found to merge asymptotically. The critical exponents
$\beta=0.355$ and $\gamma=1.07$, extracted from our data do not
correspond to any of the well-known universality classes.
Furthermore, anisotropic values of $\partial T_C/\partial p_i$ and
$\partial M_s/\partial p_i$, were found with the smallest stress
dependence for the easy direction. These results may motivate
further microscopic investigations of UIr and may be used to
develop spin fluctuation theories \cite{moriya85,takahashi06} for
the treatment of real, and generally anisotropic, systems. In the
future, the magnetization and magnetostriction of UIr might be
studied under pressure to determine the critical magnetic
properties at the onset of superconductivity. Such developments,
combined with similar studies on the itinerant ferromagnets
UGe$_2$, URhGe, and ZrZn$_2$, could be decisive for determining
what stabilizes superconductivity at a magnetic quantum
instability \cite{moriya00}.

We acknowledge useful discussions with F. Hardy and T. Schwarz.
This work was supported by the Helmholtz-Gemeinschaft through the
Virtual Institute of Research on Quantum Phase Transitions and
Project VH-NG-016.


\begin{references}


\bibitem{sarrao07} J.L. Sarrao and J.D. Thompson ,
J. Phys. Soc. Jpn. {\bf76}, 051013 (2007).

\bibitem{flouquet06}
J. Flouquet et al., C.R. Physique {\bf 7}, 22 (2006).

\bibitem{thalmeier05}
P. Thalmeier et al., {\it Frontiers in superconducting materials},
Chapter 3 (Springer, Berlin, 2005).

\bibitem{mathur98} N.D. Mathur et al., Nature {\bf 394} 39 (1998).

\bibitem{levy05} F. L\'evy et al., Science {\bf309}, 1343 (2005).

\bibitem{moriya00} T. Moriya and K. Ueda, Adv. Phys. {\bf49}, 555 (2000).

\bibitem{storey07} J.G. Storey et al., Phys. Rev. B {\bf76}, 174522 (2007).

\bibitem{akazawa04} T. Akazawa et al., J. Phys.: Condens. Matter
{\bf16}, L29 (2004).

\bibitem{bauer04} E. Bauer et al., Phys. Rev. Lett. {\bf92}, 027003 (2004).

\bibitem{dommann87} A. Dommann et al., J. Magn. Magn. Mater. {\bf67}, 323 (1987).

\bibitem{galatanu05} A. Galatanu et al., J. Phys. Soc.
Jpn. {\bf74}, 1582 (2005).

\bibitem{moriya85} T. Moriya, {\it Spin fluctuations
in itinerant electron magnetism} (Springer, Berlin, 1985).

\bibitem{sakarya07} S. Sakarya et al., J. Magn. Magn. Mater. {\bf310}, 1564 (2007).

\bibitem{saxena00} S.S. Saxena et al., Nature {\bf406}, 587 (2000).

\bibitem{huxley01} A. Huxley et al., Phys. Rev. B 63,
144519 (2001).

\bibitem{aoki01} D. Aoki et al., Nature {\bf413}, 613 (2001).

\bibitem{hardy05} F. Hardy et al., Physica B {\bf359-361}, 1111 (2005).

\bibitem{uhlarz04} M. Uhlarz et al., Phys. Rev. Lett. {\bf93}, 256404 (2004).

\bibitem{yelland05} E.A. Yelland et al., Phys. Rev. B {\bf72}, 214523 (2005).

\bibitem{moriya95} T. Moriya and T. Takimoto, J. Phys. Soc. Japan {\bf 64}, 960 (1995).

\bibitem{takahashi06} Y. Takahashi and T. Kanomata, Mater. Trans. {\bf47}, 460 (2006).

\bibitem{meingast90} C. Meingast et al., Phys. Rev. B {\bf41}, 11299, (1990).

\bibitem{pott83} R. Pott and R. Schefzyk, J. Phys. E {\bf16}, 444 (1983).

\bibitem{siratori82} K. Siratori et al., J. Phys. Soc. Jpn. {\bf51}, 2746
(1982).

\bibitem{sakarya08} S. Sakarya et al., to be published.

\bibitem{note_volume} Due to the monoclinic structure of UIr,
the approximation $\partial X/\partial p_h=\sum_{i}\partial
X/\partial p_i$, with $X=T_C$ or $M_s$ and $i=$
(10$\overline{1}$), (101), or (010), is valid within about 1 $\%$
[S. Sakarya, PhD thesis, Delft University of Technology (2006)].
Indeed, the angles between (10$\overline{1}$), (101), or (010) are
slightly different from 90 $^\circ$.

\bibitem{arrott67} A. Arrott and J.E. Noakes, Phys. Rev. Lett. {\bf19}, 786 (1967).

\bibitem{collins89} M.F. Collins, {\it Magnetic critical scattering} (Oxford
University Press, New York, 1989).

\bibitem{belov56} A similar critical power law of $\lambda(H)$ was predicted 60
years ago in K.P. Belov, Fiz. Metall. Metalloced. {\bf2}, 447
(1956), but only for the mean-field universality class.

\bibitem{kaiser88} A.B. Kaiser and P. Fulde, Phys. Rev. B {\bf37}, 5357 (1988).

\bibitem{knafo03}W. Knafo et al, J. Phys.: Condens. Matter {\bf15}, 3741 (2003).

\bibitem{monthoux08} P. Monthoux et al., Nature
{\bf450}, 1177 (2007).



\end{references}
\end{document}